\documentclass[journal]{IEEEtran}
\usepackage[latin9]{inputenc}
\usepackage{float}
\usepackage{booktabs}
\usepackage{amsmath}
\usepackage{amsthm}
\usepackage{amssymb}
\usepackage{graphicx}

\makeatletter

\providecommand{\tabularnewline}{\\}
\floatstyle{ruled}
\newfloat{algorithm}{tbp}{loa}
\providecommand{\algorithmname}{Algorithm}
\floatname{algorithm}{\protect\algorithmname}

\theoremstyle{plain}
\newtheorem{thm}{\protect\theoremname}
\theoremstyle{plain}
\newtheorem{lem}[thm]{\protect\lemmaname}

\IEEEoverridecommandlockouts
\usepackage{mathrsfs}
\usepackage{amsfonts}
\usepackage{ifpdf}
\usepackage{cite}

\usepackage{array}
\usepackage{mdwmath}
\usepackage{mdwtab}
\usepackage{eqparbox}
\usepackage{subfig}

\usepackage{algorithmic}
\usepackage{algorithm}
\usepackage{epsfig}
\usepackage{times}
\usepackage{threeparttable}
\usepackage{multirow}
\usepackage{xcolor}
\usepackage{amsfonts}\usepackage{graphicx}
\usepackage{textcomp}
\usepackage{xcolor}
\usepackage[normalem]{ulem}
\usepackage{tabularx}
\allowdisplaybreaks[4]

\def\BibTeX{{\rm B\kern-.05em{\sc i\kern-.025em b}\kern-.08em
		T\kern-.1667em\lower.7ex\hbox{E}\kern-.125emX}}

\usepackage{amsthm}
\usepackage{geometry}
\makeatother

\providecommand{\lemmaname}{Lemma}
\providecommand{\theoremname}{Theorem}

\begin{document}
\title{\noindent Task-oriented Age of Information for Remote Monitoring Systems}
\author{\noindent \IEEEauthorblockN{Shuying Gan\IEEEauthorrefmark{1}, Xijun~Wang\IEEEauthorrefmark{1},
Chao Xu\IEEEauthorrefmark{2}, and Xiang Chen\IEEEauthorrefmark{1}\\
}\IEEEauthorblockA{\IEEEauthorrefmark{1}School of Electronics and Information Engineering,
Sun Yat-sen University, Guangzhou, 510006, China\\
\IEEEauthorrefmark{2}School of Information Engineering, Northwest
A\&F University, Yangling, 712100, China\\
Email: ganshy7@mail2.sysu.edu.cn, wangxijun@mail.sysu.edu.cn, cxu@nwafu.edu.cn,
\\
chenxiang@mail.sysu.edu.cn}\thanks{This work was supported in part by the National Natural Science Foundation
of China under Grants 62271513 and 62271413, and in part by the Research
Fund under the Shaanxi Province Innovation Capability Support Program
under Grant 2023KJXX-010.}}

\maketitle

\begin{abstract}
The emergence of intelligent applications has fostered the development
of a task-oriented communication paradigm, where a comprehensive,
universal, and practical metric is crucial for unleashing the potential
of this paradigm. To this end, we introduce an innovative metric,
the Task-oriented Age of Information (TAoI), to measure whether the
content of information is relevant to the system task, thereby assisting
the system in efficiently completing designated tasks. Also, we study
the TAoI in a remote monitoring system, whose task is to identify
target images and transmit them for subsequent analysis. We formulate
the dynamic transmission problem as a Semi-Markov Decision Process
(SMDP) and transform it into an equivalent Markov Decision Process
(MDP) to minimize TAoI and find the optimal transmission policy. Furthermore,
we demonstrate that the optimal strategy is a threshold-based policy
regarding TAoI and propose a relative value iteration algorithm based
on the threshold structure to obtain the optimal transmission policy.
Finally, simulation results show the superior performance of the optimal
transmission policy compared to the baseline policies.
\end{abstract}


\section{Introduction}

Generally, conventional communications rely on Shannon's channel coding
theory to achieve reliable transmission from sources to destinations
\cite{Luo2022}. The core idea is to abstract information into bits
and design source coding/decoding, channel coding/decoding, and modulation/demodulation
parts to minimize bit/symbol error rate or signal distortion measures
(e.g., Mean Square Error) and achieve error-free replication of bits
from source to destination \cite{Sagduyu2023}. This approach has
demonstrated tremendous success across a range of voice and data communication
systems \cite{Niu2022}. However, with the emergence of intelligent
systems such as real-time cyber-physical systems, interactive systems,
and autonomous multi-agent systems, the communication goals of these
systems are no longer to reconstruct the underlying message but to
enable the destination to make the right inference or to take the
right decision at the right time and within the right context \cite{Guenduez2023}.
This shift poses new technical challenges to conventional communication
systems \cite{Yang2023}. Therefore, a new communication paradigm,
task-oriented communication or goal-oriented communication, has been
proposed as a promising solution \cite{Shi2023}.\thispagestyle{empty}

\begin{figure*}[!t]
\centering\includegraphics[width=0.8\textwidth]{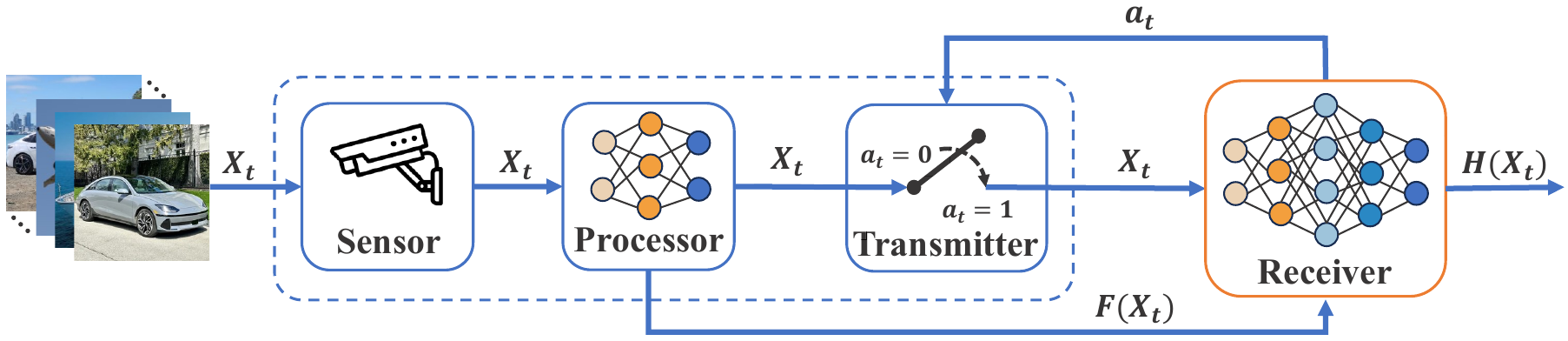}

\vspace{-0.2em}

\caption{An illustration of the task-oriented monitoring system.}

\vspace{-0.8em}

\label{Fig:System_Model}
\end{figure*}

The key to unlocking the potential of task-oriented communication
lies in a comprehensive, universal, and practical metric. This metric
measures the relevance of information to system task, thereby significantly
reducing computational and transmission costs by only acquiring, transmitting,
and reconstructing task-relevant information. In recent years, there
have been several studies on metrics for task-oriented communication
\cite{Kaul2011,Maatouk2020,Wang2022,Abolhassani2021,Nikkhah2023}.
In \cite{Kaul2011}, the Age of Information (AoI) was proposed as
a pioneering metric to capture the freshness of data perceived by
the destination. However, AoI is limited in that it cannot measure
the content of information or the dynamic changes in information content.
 To address this issue, the authors of \cite{Maatouk2020} proposed
the Age of Incorrect Information (AoII), which reflects the discrepancy
between the receiver's estimate and the actual system state. In \cite{Wang2022},
the authors introduced the Age of Changed Information (AoCI), positing
that changes in information content to be more beneficial to the system.
In \cite{Abolhassani2021}, the Age of Version (AoV) was introduced,
counting the integer difference between the versions at the database
and the local cache. The authors of \cite{Nikkhah2023} proposed the
Age of Actuation (AoA), which captures the elapsed time since the
last performed actuation at a destination based on data received by
a source.  However, none of these metrics directly assess whether
the information content is relevant to the system task.

In this paper, we consider a remote monitoring system consisting of
a sensor, a processor, a transmitter, and a receiver. Specifically,
the sensor captures real-time images, which are pre-identified by
the processor, and then the receiver decides whether to require the
transmitter to transmit the images based on the pre-identification
results and monitoring target. A new task-oriented communication metric
called Task-oriented Age of Information (TAoI) is introduced, which
directly characterizes whether the information content is relevant
to the system task. If the transmitted image matches the target, TAoI
decreases; otherwise, it increases. We focus on finding the optimal
transmission policy for the remote monitoring system to minimize TAoI.
By modeling the problem as an infinite time-horizon Semi-Markov Decision
Process (SMDP) and transforming it into an equivalent MDP with uniform
time steps, we prove that the optimal transmission policy is a threshold-type
policy. Furthermore, we propose a low-complexity relative value iteration
algorithm based on the threshold structure to obtain the optimal transmission
policy. Finally, simulation results demonstrate that the optimal transmission
policy outperforms the two baseline policies.

\section{System Overview}

\label{Sec:Section 2} 

\subsection{System Model}

As shown in Fig. \ref{Fig:System_Model}, we consider a task-oriented
monitoring system consisting of a sensor, a processor, a transmitter,
and a receiver. The system is to identify the target images and send
them to the receiver for further analysis (e.g., highway traffic analysis
and prediction). The sensor captures real-time images, which are then
pre-identified by the lightweight binary classifier in the processor.
The pre-identification result is sent to the receiver to aid in recognizing
the target image. The receiver utilizes the pre-identification result
from the processor along with the target to determine whether to instruct
the transmitter to send the image. The transmitter receives transmission
decisions from the receiver and executes the image transmission through
the channel. The receiver is equipped with a large binary classifier.
Upon receiving the image, the large binary classifier evaluates whether
it matches the target.

\begin{figure}[!t]
\centering\includegraphics[width=0.48\textwidth]{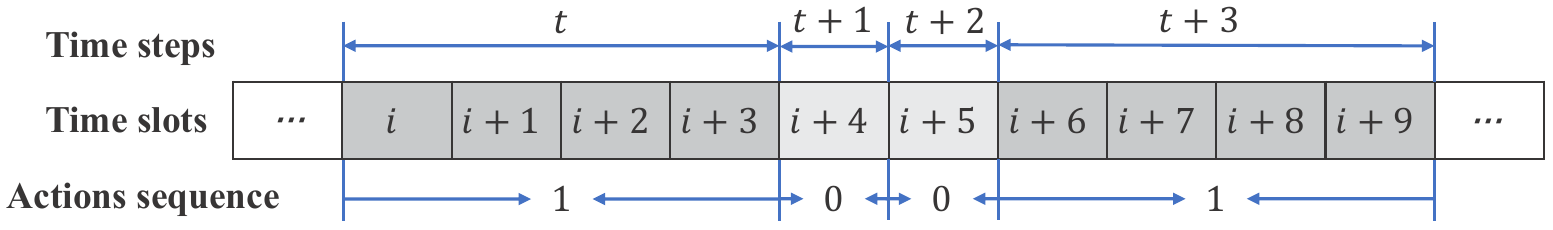}

\caption{An illustration of time slots and time steps.}

\vspace{-1em}

\label{Fig:time_slot}
\end{figure}

A time-slotted system is considered as shown in Fig. \ref{Fig:time_slot},
where the duration of each time slot is $\tau$ (in seconds) and a
decision epoch of the receiver is specified as a time step. At the
beginning of time step $t$, the sensor captures a fresh image $X_{t}\in\mathcal{X}$,
whose label is defined by $Y_{t}\in\{0,1\}$. Here, $Y_{t}=1$ indicates
that the image $X_{t}$ contains content of interest to the monitoring
task (i.e., image $X_{t}$ is the target image of the monitoring task),
and $Y_{t}=0$ indicates it does not. We define the probability that
the label $Y_{t}$ of image $X_{t}$ at time step $t$ is 1 as $\mathrm{Pr}\left(Y_{t}=1\right)=q$,
and the probability that the label $Y_{t}$ of image $X_{t}$ at time
step $t$ is 0 as $\mathrm{Pr}\left(Y_{t}=0\right)=1-q$. The processor
pre-identifies the image $X_{t}$ and sends the obtained pre-identification
result $F(X_{t})$ to the receiver, where $F(X_{t})\in\{0,1\}$. Note
that the processor may provide incorrect information. Let $p_{A}$
and $p_{B}$ respectively denote the misidentification probabilities
of image $X_{t}$ for labels $Y_{t}=0$ and $Y_{t}=1$, i.e.,

\noindent 
\begin{align}
 & p_{A}\triangleq\mathrm{Pr}(F(X_{t})=1|Y_{t}=0),\forall t,\label{Eq:misclassification_0}\\
 & p_{B}\triangleq\mathrm{Pr}(F(X_{t})=0|Y_{t}=1),\forall t.\label{Eq:misclassification_1}
\end{align}

\noindent Particularly, we denote $g$ as the probability that the
processor's 

\noindent pre-identification result for image $X_{t}$ is 1, i.e., 

\noindent 
\begin{align}
g\triangleq & \mathrm{Pr}(F(X_{t})=1)=\mathrm{Pr}(F(X_{t})=1|Y_{t}=0)\mathrm{Pr}(Y_{t}=0)\nonumber \\
 & +\mathrm{Pr}(F(X_{t})=1|Y_{t}=1)\mathrm{Pr}(Y_{t}=1)\nonumber \\
= & {p}_{A}(1-q)+(1-{p}_{B})q,\forall t.\label{Eq:pre-identify}
\end{align}

Based on the pre-identification result $F(X_{t})$ and the target,
the receiver must determine whether to request the transmitter to
transmit the image $X_{t}$. Let $a_{t}\in\left\{ 0,1\right\} $ denote
the transmission decision at time step $t$, where $a_{t}=1$ indicates
that the transmitter transmits the image $X_{t}$ to the receiver,
and $a_{t}=0$, otherwise. We assume that each image from the sensor
to the receiver takes $T_{u}$ time slots. Note that the duration
of a time step is not uniform. Specifically, let $L(a_{t})$ denote
the number of time slots in time step $t$ with action $a_{t}$ being
taken, which can be expressed as

\vspace{-0.2em}

\noindent 
\begin{align}
L(a_{t})=\begin{cases}
1, & \text{ if }a_{t}=0\\
T_{u}, & \text{ if }a_{t}=1
\end{cases} & .\label{Eq:time_step}
\end{align}

\vspace{-0.2em}

When the image $X_{t}$ arrives at the receiver, it is processed by
a large binary classifier. We assume that the classifier is perfectly
accurate, meaning the identification result $H(X_{t})$ is identical
to the image label $Y_{t}$, i.e., $H(X_{t})=Y_{t}$. We denote by
$d_{t}\in\{0,1\}$ an indicator for whether the monitoring task is
successful at time step $t$. If $d_{t}=1$, then the identification
result $H(X_{t})$ matches the target (i.e, $H(X_{t})=1$) . Otherwise,
it indicates they are different (i.e, $H(X_{t})\neq1$). In particular,
the probabilities for success and failure of the monitoring task are
given as follows:

\noindent 
\begin{align}
\mathrm{Pr}(d_{t}=1)= & \mathrm{Pr}(H(X_{t})=1)=\mathrm{Pr}(Y_{t}=1)\label{Eq:indicator_11}\\
= & {(1-\hat{p}_{A})\mathrm{Pr}(F(X_{t})=1)+\hat{p}_{B}\mathrm{Pr}(F(X_{t})=0)},\nonumber 
\end{align}

\vspace{-0.5em}

\noindent 
\begin{align}
\mathrm{Pr}(d_{t}=0)= & \mathrm{Pr}(H(X_{t})\neq1)=\mathrm{Pr}(Y_{t}\neq1)\label{Eq:indicator_00}\\
= & {\hat{p}_{A}\mathrm{Pr}(F(X_{t})=1)+(1-\hat{p}_{B})\mathrm{Pr}(F(X_{t})=0)},\nonumber 
\end{align}

\noindent where

\noindent 
\begin{align}
\hat{p}_{A} & \triangleq\mathrm{Pr}(Y_{t}=0|F(X_{t})=1)=\dfrac{(1-q)p_{A}}{(1-q)p_{A}+q(1-p_{B})},\label{Eq:misclassification_2}\\
\hat{p}_{B} & \triangleq\mathrm{Pr}(Y_{t}=1|F(X_{t})=0)=\dfrac{qp_{B}}{(1-q)(1-p_{A})+qp_{B}}.\label{Eq:misclassification_3}
\end{align}

\subsection{Task-oriented Age of Information}

In many real-time systems, AoI is extensively used to quantify the
freshness of data perceived by the receiver, thereby enhancing the
utility of decision-making processes \cite{Yates2021}. These efforts
are driven by the consensus that freshly received data typically contains
more valuable information. However, AoI does not provide a direct
measure of the data's content and ignores the relevance of information
content to system task. The metric we proposed, the TAoI, differs
from AoI in that TAoI not only captures the time lag of information
received at the destination but also assesses whether the content
of this information is relevant to the system task.

For our system, TAoI decreases only when the monitoring task is successful;
otherwise, it increases. Formally, let $U_{t}$ denote the time step
at which the receiver receives the target image. Then, the TAoI at
the $i$th time slot of time step $t$ can be defined as

\vspace{-1em}

\noindent 
\begin{align}
\Delta_{t,i}=\sum_{n=U_{t}}^{t-1}L(a_{n})+i-1 & ,\label{Eq:TAoI}
\end{align}

\vspace{-0.6em}

\noindent where the first term is the total number of time slots in
the previous time steps since $U_{t}$. For ease of exposition, we
represent the TAoI at the beginning of time step $t$ as $\Delta_{t}$.
That is, $\Delta_{t}=\Delta_{t,1}=\sum_{n=U_{t}}^{t-1}L(a_{n})$.
We let $\hat{\Delta}$ be the upper limit of the TAoI, which is assumed
to be finite but can be arbitrarily large. When the transmitter sends
an image to the receiver, and this image is the target image (i.e.,
$a_{t}=1$ and $d_{t}=1$), we specify that TAoI decreases to $T_{u}$.
If the transmitter sends an image that is not the target image (i.e.,
$a_{t}=1$ and $d_{t}=0$), then TAoI increases by $T_{u}$. When
the receiver decides not to request the transmitter to send the image
(i.e., $a_{t}=0$), TAoI increases by one. Thus, the dynamics of TAoI
can be shown as follows: 

\vspace{-0.4em}

\noindent 
\begin{align}
\Delta_{t+1}=\begin{cases}
T_{u}, & \text{ if }a_{t}=1\;\text{and}\;d_{t}=1;\\
\min\{\Delta_{t}+T_{u},\hat{\Delta}\}, & \text{ if }a_{t}=1\;\text{and}\;d_{t}=0;\\
\min\{\Delta_{t}+1,\hat{\Delta}\}, & \text{ if }a_{t}=0.
\end{cases}\label{Eq:AoI}
\end{align}

\vspace{-0.4em}

\noindent In addition, we illustrate an example of the TAoI evolution
with $T_{u}=4$ in Fig. \ref{Fig:TAoI_dynamic}.

\vspace{-1em}

\subsection{Optimization Problem}

In this paper, we aim at finding an optimal transmission policy $\pi=\left(a_{1},a_{2},\cdots\right)$
that minimizes the long-term average TAoI. Therefore, the dynamic
transmission problem can be formulated as follows: 

\vspace{-0.8em}

\noindent 
\begin{align}
\min_{\pi}\limsup_{T\rightarrow\infty}{\dfrac{\mathbb{E}\left[\sum_{t=1}^{T}\Delta_{t}\right]}{\mathbb{E}\left[\sum_{t=1}^{T}L(a_{t})\right]}} & .\label{Eq:Our_pro}
\end{align}

\vspace{-1.4em}

\section{SMDP Formulation and Solution\label{Sec:Section 3} }

\subsection{SMDP Formulation}

SMDP extends MDP to handle situations where the time

\noindent intervals between decision instants are not constant, as
occurs with the dynamic transmission problem considered here. To this
end, we formulate the problem as an infinite time-horizon SMDP problem,
which consists of a tuple $\left(\mathcal{S},\mathcal{A},t^{+},\mathrm{Pr}(\cdot,\cdot),R(\cdot,\cdot)\right)$
and is described as follows:

\begin{figure}[!t]
\centering \includegraphics[width=0.4\textwidth,height=0.2\paperheight]{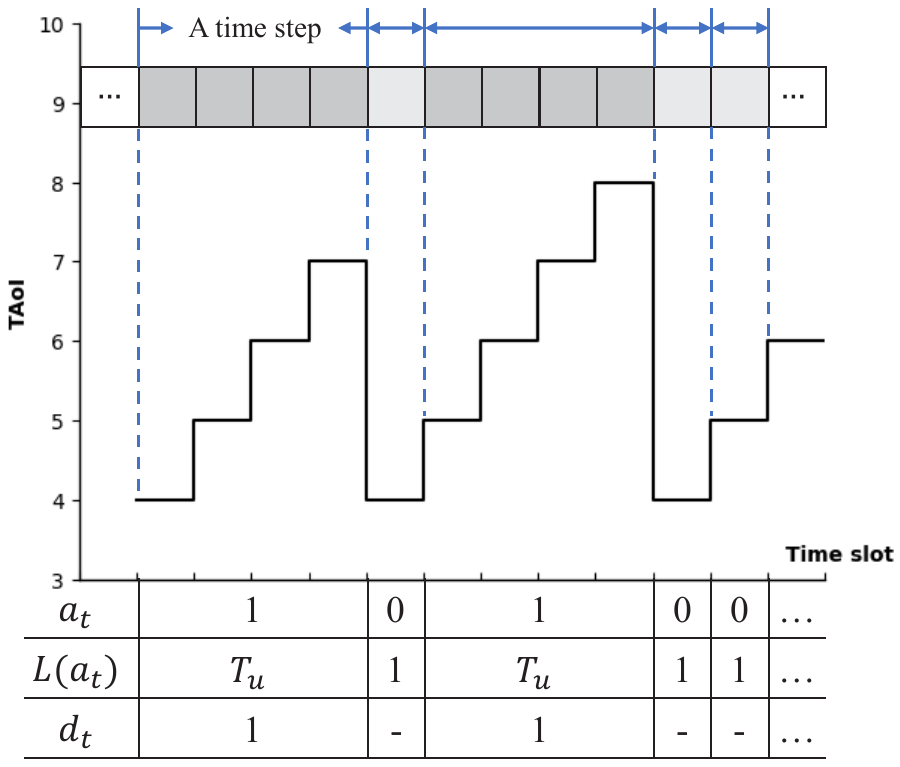}\caption{An illustration of the evolution of TAoI ($T_{u}=3$).}

\vspace{-1em}
\label{Fig:TAoI_dynamic}
\end{figure}

\textbf{1) State space $\mathcal{S}$}: The state $\mathbf{s}_{t}$
of the SMDP at time step $t$ is defined as $\mathbf{s}_{t}\triangleq(\Delta_{t},F(X_{t}))$,
where $\Delta_{t}$ denotes the TAoI at the beginning of time step
$t$ and $F(X_{t})$ represents the pre-identification result for
the image $X_{t}$ at time step $t$. The space of all possible states
is denoted by $\mathcal{S}$, and it is finite because the TAoI is
bounded by an upper limit $\hat{\Delta}$.

\textbf{2) Action space $\mathcal{A}$}: The action at time step $t$
is the transmission decision $a_{t}$ and the action space is $\mathcal{A}\triangleq\{0,1\}$.

\textbf{3) Decision epoch $t^{+}$}: A transmission decision is made
at the beginning of a time step. The time interval $L(a_{t})$ between
two adjacent decision epochs depends on the action $a_{t}$ taken
at time step $t$, as detailed in (\ref{Eq:time_step}).

\textbf{4) Transition probability $\mathrm{Pr}(\cdot,\cdot)$}: Given
current state $\mathbf{s}_{t}=(\Delta_{t},F(X_{t}))$ and action $a_{t}$,
the transition probability to the next state $\mathbf{s}_{t+1}=(\Delta_{t+1},F(X_{t+1}))$
is denoted by $\mathrm{Pr}(\mathbf{s}_{t+1}|\mathbf{s}_{t},a_{t})$.
According to the TAoI evolution dynamic in (\ref{Eq:AoI}), the transition
probability can be given in Table \ref{Ta:Tran_pro}.

\textbf{5) Reward function $R(\cdot,\cdot)$}: Let $\Delta_{t}$ be
the instantaneous reward under state $\mathbf{s}_{t}$ given action
$a_{t}$, i.e., 

\vspace{-1.6em}

\begin{align}
R(\mathbf{s}_{t},a_{t}) & =R((\Delta_{t},F(X_{t})),a_{t})\nonumber \\
 & =\sum_{i=1}^{L(a_{t})}\Delta_{t,i}=\sum_{i=1}^{L(a_{t})}\Delta_{t}+i-1\nonumber \\
 & =L(a_{t})[\Delta_{t}+\dfrac{1}{2}\left(L(a_{t})-1\right)].
\end{align}

\vspace{-0.4em}

\begin{table}[!t]
\global\long\def\arraystretch{1.5}%
\centering \caption{Transition probability}

\vspace{-0.2em}
\begin{tabular}{cccc}
\toprule 
$\mathrm{Pr}(\mathbf{s}_{t+1}|\mathbf{s}_{t},a_{t})$ & $\mathbf{s}_{t}$ & $a_{t}$ & $\mathbf{s}_{t+1}$\tabularnewline
\midrule 
$(1-\hat{p}_{A})g$ & $(\Delta_{t},1)$ & 1 & $(T_{u},1)$\tabularnewline
$\hat{p}_{B}g$ & $(\Delta_{t},0)$ & 1 & $(T_{u},1)$\tabularnewline
$(1-\hat{p}_{A})(1-g)$ & $(\Delta_{t},1)$ & 1 & $(T_{u},0)$\tabularnewline
$\hat{p}_{B}(1-g)$ & $(\Delta_{t},0)$ & 1 & $(T_{u},0)$\tabularnewline
$\hat{p}_{A}g$ & $(\Delta_{t},1)$ & 1 & $(\min\{\Delta_{t}+T_{u},\hat{\Delta}\},1)$\tabularnewline
$(1-\hat{p}_{B})g$ & $(\Delta_{t},0)$ & 1 & $(\min\{\Delta_{t}+T_{u},\hat{\Delta}\},1)$\tabularnewline
$\hat{p}_{A}(1-g)$ & $(\Delta_{t},1)$ & 1 & $(\min\{\Delta_{t}+T_{u},\hat{\Delta}\},0)$\tabularnewline
$(1-\hat{p}_{B})(1-g)$ & $(\Delta_{t},0)$ & 1 & $(\min\{\Delta_{t}+T_{u},\hat{\Delta}\},0)$\tabularnewline
$g$ & $(\Delta_{t},1)$ & 0 & $(\min\{\Delta_{t}+1,\hat{\Delta}\},1)$\tabularnewline
$g$ & $(\Delta_{t},0)$ & 0 & $(\min\{\Delta_{t}+1,\hat{\Delta}\},1)$\tabularnewline
$1-g$ & $(\Delta_{t},1)$ & 0 & $(\min\{\Delta_{t}+1,\hat{\Delta}\},0)$\tabularnewline
$1-g$ & $(\Delta_{t},0)$ & 0 & $(\min\{\Delta_{t}+1,\hat{\Delta}\},0)$\tabularnewline
\bottomrule
\end{tabular}\label{Ta:Tran_pro}

\vspace{-1em}
\end{table}

Given an initial system state $\mathbf{s}_{1}$, the dynamic transmission
problem (\ref{Eq:Our_pro}) can be expressed as

\noindent \vspace{-1.4em}
\begin{align}
\min_{\pi}\limsup_{T\rightarrow\infty}\dfrac{\mathbb{E}\left[\sum_{t=1}^{T}R(\mathbf{s}_{t},a_{t})\mid\mathbf{s}_{1}\right]}{\mathbb{E}\left[\sum_{t=1}^{T}L(a_{t})\right]} & .\label{Eq:trans_policy}
\end{align}

\noindent Due to the non-uniform duration of time steps, the average
reward in (\ref{Eq:trans_policy}) is defined as the limit of the
expected total reward over a finite number of time steps divided by
the expected cumulative time of these time steps. In this paper, our
objective is to find a stationary deterministic optimal transmission
policy that solves the long-term average TAoI minimization problem
as (\ref{Eq:trans_policy}). To this end, we initially apply uniformization
to transform the SMDP into an equivalent discrete-time MDP \cite{Puterman2014,Tijms2003}.
Denoting the state and action spaces of the transformed MDP as $\hat{\mathcal{S}}$
and $\hat{\mathcal{A}}$ respectively, they remain identical to those
in the original SMDP, i.e., $\hat{\mathcal{S}}=\mathcal{S}$ and $\hat{\mathcal{A}}=\mathcal{A}$.
For any $\mathbf{s}=(\Delta,F(X))\in\hat{\mathcal{S}}$ and $a\in\hat{\mathcal{A}}$,
the reward in the MDP can be given by

\vspace{-0.2em}

\noindent 
\begin{align}
\bar{R}((\Delta,F(X)),a)=\Delta+\dfrac{1}{2}(L(a)-1) & ,\label{Eq:reward}
\end{align}

\noindent \vspace{-0.2em}
and the transition probability is given by 

\noindent \vspace{-1em}
\begin{align}
\bar{p}(\mathbf{s}^{\prime}|\mathbf{s},a)=\begin{cases}
\frac{\epsilon}{L(a)}p(\mathbf{s}^{\prime}|\mathbf{s},a), & \mathbf{s}^{\prime}\neq\mathbf{s}\\
1-\frac{\epsilon}{L(a)}, & \mathbf{s}^{\prime}=\mathbf{s}
\end{cases} & ,\label{Eq:Pro}
\end{align}

\noindent \vspace{-0.2em}
where $\epsilon$ is chosen in $(0,\min_{a}L(a)]$. Next, the average
TAoI under policy $\pi$ is given by

\vspace{-0.2em}

\noindent 
\begin{align}
V_{\pi}(\mathbf{s})=\dfrac{1}{T}\limsup_{T\rightarrow\infty}{\mathbb{E}\left[\sum_{t=1}^{T}\bar{R}(\mathbf{s}_{t},a_{t})\mid\mathbf{s}_{1}\right]} & .\label{Eq:V}
\end{align}
For an infinite horizon, we focus on the set of deterministic stationary
policies $\Pi$, where $\pi=\{a_{1},a_{2},\cdots\}\in\Pi$ such that
$a_{t_{1}}=a_{t_{2}}$ when $\mathbf{s}_{t_{1}}=\mathbf{s}_{t_{2}}$
for any ${t_{1}},{t_{2}}$. Thus, we omit the time index in the sequel.
The objective is to find a policy $\pi\in\Pi$ that minimizes the
average TAoI. Specifically, if there exists a policy that minimizes
(\ref{Eq:V}), i.e., 

\noindent 
\begin{align}
\min_{\pi\in\Pi}V_{\pi}(\mathbf{s}) & ,\label{Eq:opt_pi}
\end{align}

\noindent we refer to this policy as the average TAoI optimal policy
and denote it as $\pi^{*}$. 

According to \cite[Theorem 8.4.5]{Puterman2014}, a deterministic
stationary average optimal policy exists for the finite-state finite-action
average-reward MDP if the reward function is bounded and the MDP is
unichain. Therefore, before analyzing the stationary deterministic
optimal policy for average TAoI, we need to prove the existence of
such a policy. We then examine the two conditions for the existence
of a deterministic stationary policy. Firstly, the reward of the MDP
is bounded, as the instantaneous reward is defined by the TAoI. Secondly,
since the state $(\hat{\Delta},F(X))$ $\text{(}F(X)\in\{0,1\})$
is reachable from all other states, and $(\hat{\Delta},0)$ and $(\hat{\Delta},1)$
can reach each other through other states, the induced Markov chain
has a single recurrent class. Consequently, the MDP is unichain. Therefore,
a stationary and deterministic optimal policy exists.

We can obtain the optimal policy $\pi^{*}$ for the original SMDP
by solving the Bellman equation in (\ref{Eq:Bellman}). According
to \cite{Bertsekas2012}, we have

\vspace{-0.4em}

\noindent 
\begin{align}
V^{*}+h(\mathbf{s})=\min_{a\in\mathcal{A}}\left\{ \bar{R}(\mathbf{s},a)+\sum_{\mathbf{s}^{\prime}\in\mathcal{S}}\bar{p}(\mathbf{s}^{\prime}|\mathbf{s},a)h(\mathbf{s}^{\prime})\right\} ,\;\forall\mathbf{s}\in\mathcal{S} & ,\label{Eq:Bellman}
\end{align}

\vspace{-0.4em}

\noindent where $V^{*}$ represents the optimal value to (\ref{Eq:trans_policy})
for all initial states, and $h(\mathbf{s})$ is the relative value
function for the discrete-time MDP. Similarly, the state-action value
function can be obtained as follows:

\vspace{-0.4em}

\noindent 
\begin{align}
Q(\mathbf{s},a)=\bar{R}(\mathbf{s},a)+\sum_{\mathbf{s}^{\prime}\in\mathcal{S}}\bar{p}(\mathbf{s}^{\prime}|\mathbf{s},a)h(\mathbf{s}^{\prime}),\;\forall\mathbf{s}\in\mathcal{S},a & \in\mathcal{A}\text{.}\label{Eq:Bellman-1}
\end{align}

\vspace{-0.4em}

\noindent Then, we have

\vspace{-0.4em}

\noindent 
\begin{align}
V^{*}+h(\mathbf{s})=\min_{a\in\mathcal{A}}Q(\mathbf{s},a) & .\label{Eq:opt_polic-1}
\end{align}

\vspace{-0.4em}

\noindent Thus, the optimal policy $\pi^{*}$ for any $\mathbf{s}\in\mathcal{S}$
can be given by 

\noindent \vspace{-0.4em}
\begin{align}
\pi^{*}(\mathbf{s})=\arg\min_{a\in\mathcal{A}}Q(\mathbf{s},a) & .\label{Eq:opt_polic}
\end{align}

\vspace{-1em}

\subsection{Structural Analysis and Optimal Transmission Policy}

To solve (\ref{Eq:opt_polic}), we use the Relative Value Iteration
(RVI) algorithm to obtain the optimal policy. Firstly, we set any
arbitrary but fixed reference state $\mathbf{s^{\dagger}}=(\Delta^{\dagger},F^{\dagger}(X))$,
and randomly initialize $V_{0}(\mathbf{s})$ and $h_{0}(\mathbf{s})$.
Then, each iteration of the RVI algorithm is as follows:

\vspace{-0.4em}

\noindent 
\begin{align}
Q_{k+1}(\mathbf{s},a)\leftarrow\bar{R}(\mathbf{s},a)+\sum_{\mathbf{s}^{\prime}\in\mathcal{S}}\bar{p}(\mathbf{s}^{\prime}|\mathbf{s},a)h_{k}(\mathbf{s}^{\prime}) & ,\label{Eq:RVI-1}
\end{align}

\vspace{-0.5em}

\noindent 
\begin{align}
V_{k+1}(\mathbf{s})\leftarrow\min_{a\in\mathcal{A}}Q_{k+1}(\mathbf{s},a) & ,\label{Eq:RVI-2}
\end{align}

\vspace{-0.5em}

\noindent 
\begin{align}
h_{k+1}(\mathbf{s})\leftarrow V_{k+1}(\mathbf{s})-V_{k+1}(\mathbf{s^{\dagger}}) & \text{.}\label{Eq:RVI-3}
\end{align}
When the above algorithm iterates until convergence of the value function,
the optimal policy corresponding to each state can be derived from
(\ref{Eq:opt_polic}). From (\ref{Eq:RVI-2}), it is known that when
the state value function $V(s)$ is updated in each iteration, the
minimum of the state-action value function $Q$ under all actions
must be taken. Analyzing the threshold structure of the optimal policy
can effectively reduce the complexity of the value iteration algorithm.
Therefore, we prove that the optimal policy for the problem (\ref{Eq:Our_pro})
exists a threshold structure with respect to $\Delta$, as detailed
below.

To begin with, we introduce several essential properties of the value
function $V(\mathbf{s})$ (i.e., $V(\Delta,F(X))$) in the lemmas.

\vspace{-0.4em}

\begin{lem}
\label{LM:Lemma1}The value function $V(\Delta,F(X))$ is non-decreasing
with $\Delta$ for any given $F(X)$.
\end{lem}
\vspace{-0.4em}

\begin{IEEEproof}
See Section II-A in the online supplemental materials \cite{Supplementary}.
\end{IEEEproof}
\vspace{-0.4em}

\begin{lem}
\label{LM:Lemma2} Given $F(X)$, the value function $V(\Delta,F(X))$
is concave in $\Delta$. 
\end{lem}
\vspace{-0.4em}

\begin{IEEEproof}
See Section II-B in the online supplemental materials \cite{Supplementary}.
\end{IEEEproof}
Since the value function $V(\Delta,F(X))$ is concave, its slope does
not increases. The lower bound of the slope of $V(\Delta,F(X))$ is
given by the following lemma.

\vspace{-0.4em}

\begin{lem}
\label{LM:Lemma3} For any $\mathbf{s}_{1}=(\Delta_{1},F(X))$, $\mathbf{s}_{2}=(\Delta_{2},F(X))\in\mathcal{S}$
with $\Delta_{1}\leq\Delta_{2}$, we have $V_{k}(\Delta_{2},F(X))-V_{k}(\Delta_{1},F(X))\geq\dfrac{L(a)}{\epsilon(1-p_{1})}(\Delta_{2}-\Delta_{1})$,
where $p_{1}=\hat{p}_{A}$ if $F(X)=1$ and $p_{1}=1-\hat{p}_{B}$
if $F(X)=0$.
\end{lem}
\vspace{-0.4em}

\begin{IEEEproof}
See Section II-C in the online supplemental materials \cite{Supplementary}.
\end{IEEEproof}
Based on the above lemmas, we can derive the structure of the optimal
transmission policy, as stated in the theorem.

\vspace{-0.4em}

\begin{thm}
\label{Th:Theorem1} Given $F(X)$, there exists a stationary deterministic
optimal policy that is of threshold-type in $\Delta$. Specifically,
if $\Delta\geq\Omega$, the $\pi^{*}=1$, where $\Omega$ denotes
the threshold given pair of $\Delta$ and $F(X)$. 
\end{thm}
\vspace{-0.4em}

\begin{IEEEproof}
See Appendix \ref{PR:proof_Theorem1-1}. 
\end{IEEEproof}
According to Theorem 1, a threshold $\Omega$ exists within the optimal
transmission policy that can be applied to the RVI algorithm to reduce
the computational complexity. Specifically, if an optimal policy exhibits
a threshold structure with respect to $\Delta$, then the optimal
policy can be directly obtained in lines 5 and 6 of Algorithm \ref{alg:algorithm1},
without the need to first calculate the state-action value function
under different actions and then select the optimal action. Based
on the optimal policy, the state value function $\ensuremath{V}$
and the relative value function $h$ can be directly updated, as shown
in lines 13 to 15 of Algorithm \ref{alg:algorithm1}. Further details
are provided in Algorithm \ref{alg:algorithm1}.

\begin{algorithm}[t!]
\caption{Relative Value Iteration Based on the Threshold Structure}
\label{alg:algorithm1} \begin{algorithmic}[1]

\STATE \textbf{Initialization:} Set $V(\mathbf{s})=0$, $h(\mathbf{s})=0$
and $h^{'}(\mathbf{s})=\infty$ for all $\mathbf{s}\in S$, select
a reference state $\mathbf{s^{\dagger}}$ , and set $\lambda=0$ and
$\bar{\lambda}$.

\REPEAT

\STATE For all $\mathbf{s}=\{\Delta,F(X)\}\in S$, $\Omega=\infty$;

\FOR {$\mathbf{s}\in S$}

\IF{$\text{\ensuremath{\Delta}}\geq\Omega$}

\STATE $\pi^{*}=1$;

\ELSE

\STATE $\pi^{*}=\arg\min\limits _{a\in\mathcal{A}}\left\{ \bar{R}(\mathbf{s},\pi^{*})+\sum\limits _{\mathbf{s}^{\prime}\in\mathcal{S}}\bar{p}(\mathbf{s}^{\prime}|\mathbf{s},\pi^{*})h(\mathbf{s}^{\prime})\right\} $;

\IF{$\pi^{*}=1$}

\STATE $\Omega=\Delta$;

\ENDIF

\ENDIF

\STATE $V(\mathbf{s})=\bar{R}(\mathbf{s},\pi^{*})+\sum\limits _{\mathbf{s}^{\prime}\in\mathcal{S}}\bar{p}(\mathbf{s}^{\prime}|\mathbf{s},\pi^{*})h(\mathbf{s}^{\prime})$
;

\STATE $h^{'}(\mathbf{s})=h(\mathbf{s})$;

\STATE $h(\mathbf{s})=V(\mathbf{s})-V(\mathbf{s^{\dagger}})$;

\STATE $\lambda=\text{{max}(\ensuremath{\lambda,}\ensuremath{\mid h(\mathbf{s})-h^{'}(\mathbf{s})\mid}}$);

\ENDFOR

\UNTIL{$\lambda$\textless$\bar{\lambda}$}

\STATE \textbf{Output:} The optimal transmission policy $\pi^{*}$.

\end{algorithmic}
\end{algorithm}

\section{Simulation results}

\label{Sec:Section 4} 

In this section, we conduct simulations to verify the feasibility
and effectiveness of the optimal transmission policy. We consider
two baseline policies: the always transmit policy and the pre-identification
based policy. In the always transmit policy, regardless of the current
state (i.e., TAoI and pre-identification result), the receiver requests
transmission from the transmitter. In the pre-identification based
policy, the receiver completely relies on the processor and makes
transmission decisions based on the pre-identification result.

\begin{figure}[!t]
\centering\includegraphics[width=0.42\textwidth]{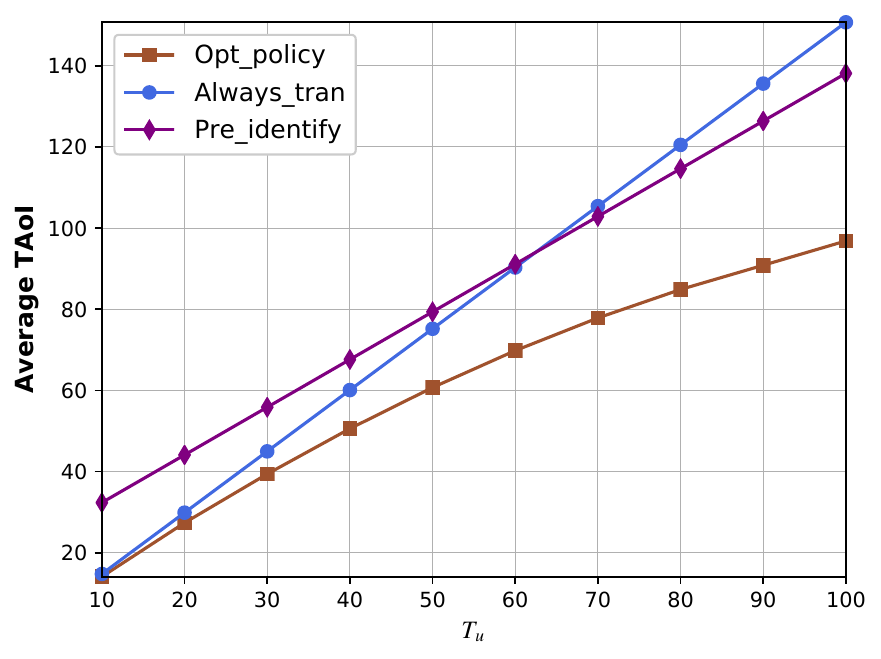}

\vspace{-0.4em}

\caption{Average TAoI versus $T_{u}$ ($q=0.9$, $p_{A}=p_{B}=0.1$).}

\vspace{-1em}

\label{Fig:avg_AoI_Pronew09_Pre1_T_10_100}
\end{figure}

In Fig. \ref{Fig:avg_AoI_Pronew09_Pre1_T_10_100}, the average TAoI
of the optimal transmission policy and the two policies are compared
with respect to $T_{u}$. From Fig. \ref{Fig:avg_AoI_Pronew09_Pre1_T_10_100},
it is observed that as $T_{u}$ increases, the gap between the optimal
transmission policy and the other two baseline policies increases.
This is because as $T_{u}$ increases, the cost of task failure becomes
larger, making the advantage of the optimal policy more significant. Furthermore,
it can be inferred that when $T_{u}$ is very small, the optimal transmission
policy approaches the always transmit policy, becoming exactly the
always transmit policy when $T_{u}$ equals $1$.

\begin{figure}[!t]
\centering\includegraphics[width=0.42\textwidth]{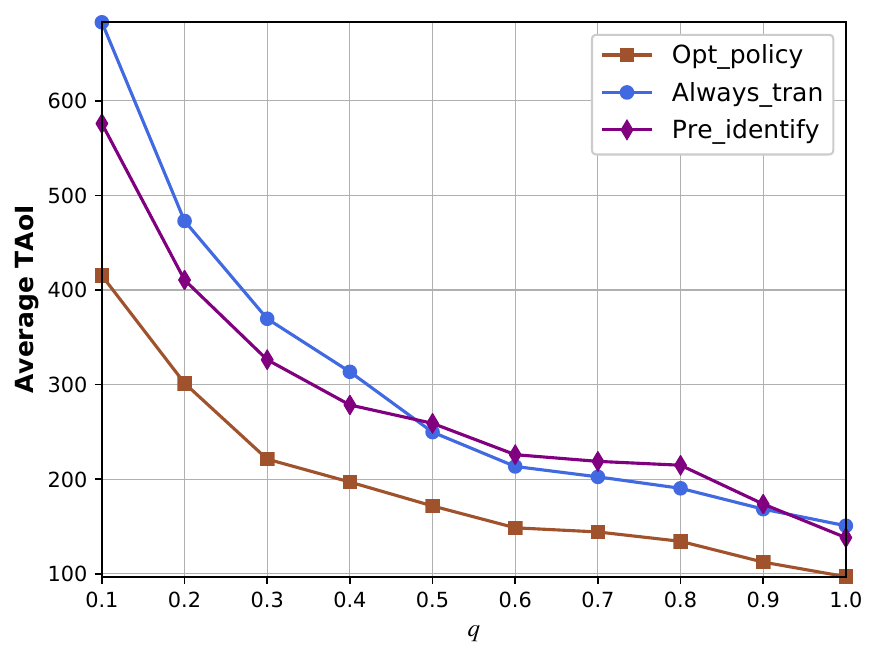}

\vspace{-0.4em}

\caption{Average TAoI versus $q$ ($T_{u}=100$, $p_{A}=p_{B}=0.1$).}

\vspace{-1em}

\label{Fig:avg_AoI_T100_Pre1_Pronew}
\end{figure}

In Fig. \ref{Fig:avg_AoI_T100_Pre1_Pronew}, the average TAoI of the
optimal transmission policy and the two policies are compared with
respect to $q$. Initially, it is noted that as $q$ increases, TAoI
decreases. This occurs because the higher frequency of the target's
appearance increases the number of successful monitoring tasks, leading
to a decrease in TAoI. Moreover, combining Fig. \ref{Fig:avg_AoI_Pronew09_Pre1_T_10_100}
and Fig. \ref{Fig:avg_AoI_T100_Pre1_Pronew}, we find that regardless
of whether $T_{u}$ and $q$ are high or low, the average TAoI of
the optimal transmission policy is lower than the average TAoI of
all baseline policies. This validates the effectiveness of the optimal
transmission policy.

\noindent 

\noindent 
\vspace{-2em}

\section{Conclusions}

\label{Sec:Section 5} 

In this paper, we introduced a new task-oriented communication metric
named TAoI, which directly measures the relevance of information content
to the system task, thereby enhancing the system's decision-making
utility. By minimizing TAoI, we explored the optimal transmission
strategy for a remote monitoring system. First, we modeled the dynamic
transmission problem as an SMDP over an infinite time horizon and
converted it into an equivalent MDP with uniform time steps. Then,
we demonstrated that the optimal transmission strategy was a threshold
policy and proposed a low-complexity threshold-structured relative
value iteration algorithm to obtain the optimal transmission policy.
Finally, simulation results showed that the optimal transmission policy
significantly outperformed baseline policies.

\appendix{}

\subsection{Proof of Theorem 1}

\label{PR:proof_Theorem1-1} For any $\mathbf{s}_{1}=(\Delta_{1},F(X))$,
$\mathbf{s}_{2}=(\Delta_{2},F(X))\in\mathcal{S}$, such that $\Delta_{1}\leq\Delta_{2}$,
we have 

\vspace{-1.4em}

\begin{align}
 & Q_{k}(\mathbf{s}_{2},a)-Q_{k}(\mathbf{s}_{1},a)-(V_{k}(\mathbf{s}_{2})-V_{k}(\mathbf{s}_{1}))\nonumber \\
= & \Delta_{2}-\Delta_{1}-\frac{\epsilon}{L(a)}(V(\Delta_{2},F(X))-V(\Delta_{1},F(X)))+\nonumber \\
 & \frac{\epsilon}{L(a)}p_{1}(V(\Delta_{2}+L(a),F(X))-V(\Delta_{1}+L(a),F(X))).
\end{align}

\noindent Since the concavity of $V(\mathbf{s})$ have been proved
in Lemma 2, we can easily see that $V(\Delta_{2}+L(a),F(X))-V(\Delta_{1}+L(a),F(X))\leq V(\Delta_{2},F(X))-V(\Delta_{1}+L(a),F(X))$.
Therefore, we have 

\vspace{-0.4em}

\noindent 
\begin{align}
 & Q_{k}(\mathbf{s}_{2},a)-Q_{k}(\mathbf{s}_{1},a)-(V_{k}(\mathbf{s}_{2})-V_{k}(\mathbf{s}_{1}))\nonumber \\
\leq & \Delta_{2}-\Delta_{1}-\frac{\epsilon}{L(a)}(V(\Delta_{2},F(X))-V(\Delta_{1}),F(X))\nonumber \\
 & +\frac{\epsilon}{L(a)}p_{1}(V(\Delta_{2},F(X))-V(\Delta_{1},F(X)))\nonumber \\
= & \Delta_{2}-\Delta_{1}-\frac{\epsilon}{L(a)}(1-p_{1})(V(\mathbf{s}_{2})-V(\mathbf{s}_{1})).
\end{align}

\noindent As proved in Lemma 3 that $V_{k}(\Delta_{2},F(X))-V_{k}(\Delta_{1},F(X))\geq[{L(a)}/{\epsilon(1-p_{1})}](\Delta_{2}-\Delta_{1})$,
it is easy to see that $Q_{k}(\mathbf{s}_{2},a)-Q_{k}(\mathbf{s}_{1},a)-(V(\mathbf{s}_{2})-V(\mathbf{s}_{1}))\leq0$. 

Now, we can prove the threshold structure of the optimal policy. Suppose
$\Delta_{2}\geq\Delta_{1}$ and $\pi^{*}(\Delta_{1},F(X))=a$, it
is easily to see that $V(\Delta_{1},F(X))=Q((\Delta_{1},F(X)),a)$,
i.e., $V(\mathbf{s}_{1})=Q(\mathbf{s}_{1},a)$. According to Theorem
1, we know that $V(\mathbf{s}_{2})-V(\mathbf{s}_{1})\geq Q(\mathbf{s}_{2},a)-Q(\mathbf{s}_{1},a)$.
Therefore, we have $V(\mathbf{s}_{2})\geq Q(\mathbf{s}_{2},a)$. Since
the value function is a minimum of two state-action cost functions,
we have $V(\mathbf{s}_{2})\leq Q(\mathbf{s}_{2},a)$. Altogether,
we can assert that $V(\mathbf{s}_{2})=Q(\mathbf{s}_{2},a)$ and $\pi^{*}(\Delta_{2},F(X))=a$.

This completes the proof of Theorem 1.

 \bibliographystyle{IEEEtran}
\bibliography{IEEEabrv,Privacy_offloading_Ref}

\begin{thebibliography}{10}
\providecommand{\url}[1]{#1}
\csname url@samestyle\endcsname
\providecommand{\newblock}{\relax}
\providecommand{\bibinfo}[2]{#2}
\providecommand{\BIBentrySTDinterwordspacing}{\spaceskip=0pt\relax}
\providecommand{\BIBentryALTinterwordstretchfactor}{4}
\providecommand{\BIBentryALTinterwordspacing}{\spaceskip=\fontdimen2\font plus
\BIBentryALTinterwordstretchfactor\fontdimen3\font minus
  \fontdimen4\font\relax}
\providecommand{\BIBforeignlanguage}[2]{{%
\expandafter\ifx\csname l@#1\endcsname\relax
\typeout{** WARNING: IEEEtran.bst: No hyphenation pattern has been}%
\typeout{** loaded for the language `#1'. Using the pattern for}%
\typeout{** the default language instead.}%
\else
\language=\csname l@#1\endcsname
\fi
#2}}
\providecommand{\BIBdecl}{\relax}
\BIBdecl

\bibitem{Luo2022}
X.~Luo, H.-H. Chen, and Q.~Guo, ``{Semantic Communications: Overview, Open
  Issues, and Future Research Directions},'' \emph{IEEE Wireless Commun.},
  vol.~29, no.~1, pp. 210--219, 2022.

\bibitem{Sagduyu2023}
Y.~E. Sagduyu, S.~Ulukus, and A.~Yener, ``{Task-Oriented Communications for
  NextG: End-to-end Deep Learning and AI Security Aspects},'' \emph{IEEE
  Wireless Commun.}, vol.~30, no.~3, pp. 52--60, 2023.

\bibitem{Niu2022}
K.~Niu, J.~Dai, S.~Yao, S.~Wang, Z.~Si, X.~Qin, and P.~Zhang, ``{A Paradigm
  Shift toward Semantic Communications},'' \emph{IEEE Commun. Mag.}, vol.~60,
  no.~11, pp. 113--119, 2022.

\bibitem{Guenduez2023}
D.~G{\"u}nd{\"u}z, Z.~Qin, I.~E. Aguerri, H.~S. Dhillon, Z.~Yang, A.~Yener,
  K.~K. Wong, and C.-B. Chae, ``{Beyond Transmitting Bits: Context, Semantics,
  and Task-Oriented Communications},'' \emph{IEEE J. Sel. Areas Commun.},
  vol.~41, no.~1, pp. 5--41, 2023.

\bibitem{Yang2023}
W.~Yang, H.~Du, Z.~Q. Liew, W.~Y.~B. Lim, Z.~Xiong, D.~Niyato, X.~Chi, X.~Shen,
  and C.~Miao, ``{Semantic Communications for Future Internet: Fundamentals,
  Applications, and Challenges},'' \emph{IEEE Commun. Surveys Tuts.}, vol.~25,
  no.~1, pp. 213--250, 2023.

\bibitem{Shi2023}
Y.~Shi, Y.~Zhou, D.~Wen, Y.~Wu, C.~Jiang, and K.~B. Letaief, ``{Task-Oriented
  Communications for 6G: Vision, Principles, and Technologies},'' \emph{IEEE
  Wireless Commun.}, vol.~30, no.~3, pp. 78--85, 2023.

\bibitem{Kaul2011}
S.~Kaul, M.~Gruteser, V.~Rai, and J.~Kenney, ``{Minimizing age of information
  in vehicular networks},'' in \emph{Proc. IEEE SECON}, 2011, pp. 350--358.

\bibitem{Maatouk2020}
A.~Maatouk, S.~Kriouile, M.~Assaad, and A.~Ephremides, ``{The Age of Incorrect
  Information: A New Performance Metric for Status Updates},'' \emph{IEEE/ACM
  Trans. Netw.}, vol.~28, no.~5, pp. 2215--2228, 2020.

\bibitem{Wang2022}
X.~Wang, W.~Lin, C.~Xu, X.~Sun, and X.~Chen, ``{Age of Changed Information:
  Content-Aware Status Updating in the Internet of Things},'' \emph{IEEE Trans.
  Wireless Commun.}, vol.~70, no.~1, pp. 578--591, 2022.

\bibitem{Abolhassani2021}
B.~Abolhassani, J.~Tadrous, A.~Eryilmaz, and E.~Yeh, ``{Fresh Caching for
  Dynamic Content},'' in \emph{Proc. IEEE INFOCOM}, Vancouver, BC, Canada, May
  2021, pp. 1--10.

\bibitem{Nikkhah2023}
A.~Nikkhah, A.~Ephremides, and N.~Pappas, ``{Age of Actuation in a Wireless
  Power Transfer System},'' in \emph{Proc. IEEE INFOCOM WKSHPS}, Hoboken, NJ,
  USA, May 2023, pp. 1--6.

\bibitem{Yates2021}
R.~D. Yates, Y.~Sun, D.~R. Brown, S.~K. Kaul, E.~Modiano, and S.~Ulukus, ``{Age
  of Information: An Introduction and Survey},'' \emph{IEEE J. Sel. Areas
  Commun.}, vol.~39, no.~5, pp. 1183--1210, 2021.

\bibitem{Puterman2014}
M.~L. Puterman, \emph{{Markov Decision Processes: Discrete Stochastic Dynamic
  Programming} (Wiley Series in Probability and Statistics)}.\hskip 1em plus
  0.5em minus 0.4em\relax Hoboken, NJ, USA: Wiley, 2005.

\bibitem{Tijms2003}
H.~C. Tijms, \emph{{A First Course in Stochastic Models}}.\hskip 1em plus 0.5em
  minus 0.4em\relax Chichester, U.K.: Wiley, Dec. 2004.

\bibitem{Bertsekas2012}
P.~Bertsekas, Dimitri, \emph{{Dynamic Programming and Optimal Control-II}},
  3rd~ed.\hskip 1em plus 0.5em minus 0.4em\relax Belmont, MA, USA: Athena Sci.,
  2007, vol.~2.

\bibitem{Supplementary}
\BIBentryALTinterwordspacing
S.~Gan, X.~Wang, C.~Xu, and X.~Chen, ``{Supplementary Materials of TAoI for
  Monitoring Systems}.'' [Online]. Available:
  \url{https://github.com/ganshuying/TAoI/blob/master/Supplementary-Materials-of-TAoI.pdf}
\BIBentrySTDinterwordspacing

\end{thebibliography}

\end{document}